# Newtonian mechanics & gravity fully model disk galaxy rotation curves without dark matter#

**Dilip G. Banhatti***

School of Physics, Madurai-Kamaraj University, Madurai 625021, India

(*Currently visiting Solid State Theory group at University of Münster, Germany)

[#For IAU Symposium 254 on Galaxy Disk in Cosmological Context, 9-14 June 2008, Copenhagen, Denmark]

**Abstract.** EGRET gamma-ray archival data used with GALPROP software show two ringlike structures in Milky Way Plane which roughly tally with distribution of stars (**[1]** & references therein). To understand fully the implications of this and similar results on detailed structure and rotation curve of especially Milky Way Disk as well as rotation curves of other galaxies as derived from spatially resolved spectroscopic data-cubes, a re-examination of the basis of the connection between mass density and rotation curve is warranted. Kenneth F Nicholson's approach **[2]**, which uses only Newtonian dynamics & gravity, is presented.

**Assumptions.** The following assumptions are made in this approach.
1. Axisymmetry (that is, azimuthal symmetry relative to the dominant structural axis) in the disk plane, taken to be the xy-plane and bilateral symmetry at every radius r in the normal direction, taken to be the z-direction.
2. The measured (that is reduced from the spatially resolved spectroscopic data cube) rotation speeds are taken to be applicable at the central (disk) plane (z = 0).
3. A definite maximum radius rmax beyond which (circular) speeds are not specified.
4. Thickness (h) variation with radius (r) is estimated at 1.491 times that of stars as roughly measured from edge-on galaxy images (like in Figure 1). This is specific to each galaxy, unless it is not measurable. Then it is taken from some other similar galaxy (after necessary normalization and scaling). The function h(r) also extends to rmax. All calculations and plots essentially apply only upto r = rmax.
5. Volume mass density rho = surface mass density / h [that is, rho(r) = SMD(r)/h(r)].

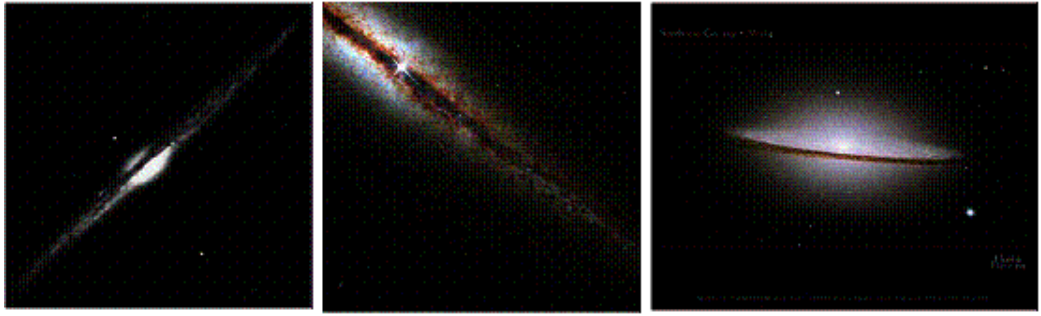

**Figure 1.** Examples of edge-on disk galaxies – NGC4565 from Jeff MacQuarrie's homepage, NGC4013 from Hubble (only half the disk is shown), and M101 (Sombrero) from Hubble.

---

**Method.** The method consists of two parts – a forward part and a reverse part. The forward part computes the rotation profile from a given mass distribution. The reverse part finds the mass distribution for a given rotation profile. Equations and coding for the forward part are checked against simple analytical solutions, giving a high degree of confidence in its use. The reverse part is done by repeated trials and feedback corrections with the forward part, to adjust the mass distribution until errors between computed and measured rotation speeds are all very small at each radius (that is, each ring out of upto 100 rings – as few as 4 or 5 rings being sometimes enough). To check the reverse part, a given mass distribution is used to find the rotation profile with the forward part, and that profile is then used as inputs for the reverse part to check against the initially input mass distribution. Thus the same degree of confidence associated with the forward part carries over to the reverse part.

**Geometry and computational variables.** Figure 2 and the following description of the computation is essentially reproduced from Kenneth F Nicholson's paper arXiv:astro-ph/0309762v1, listed in **[2]**, the other papers giving various other details and applications.

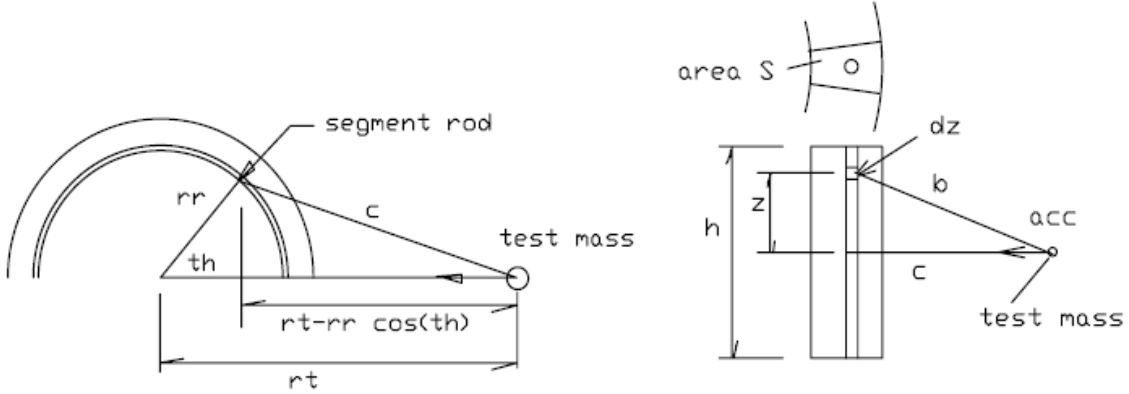

Figure 2. Disk geometry and fundamental segment

---

**Forward part**

The rotation speed of a test mass at radius rt is found by integrating the gravity effects on the test mass of all particles in the envelope:

$$v^2 = (pytks)^2 \times rt \times \int_{vol} \frac{rhoa \; dvol}{|\vec{rt}-\vec{r}|^2} \cos(phi) \, , \; kms/sec \qquad (1)$$

where $dvol$ = a volume increment, $pc^3$
$\cos(phi)$ = fraction of gravity effect caused by dvol acting toward galaxy center
$pytks$ = constant changing pcs/yr to kms/sec
$rhoa$ = actual local density, $msuns/pc^3$
$rt$ = test-mass radius, pcs
$\vec{rt}-\vec{r}$ = vector from test mass to dvol, pcs

An equivalent constant-density thickness is chosen for rings at a given radius to make the integration in z analytic. This results in a simple answer for the gravity effects of a fundamental segment of mass ddm = rho h S, where S = r dth dr. When computing is done, the equivalent thickness and density is used to find the correct local density distribution in z.

Representing the fundamental segment as a rod at the centroid* of S, the acceleration on the test mass caused by and toward the rod is found by integration to be:

$$acc = \frac{G \; ddm}{c \sqrt{c^2 + (h/2)^2}}$$

Noting that ddm = rho h rdth dr, the integral for $v^2$ at radius rt becomes:

$$v^2 = (pytks)^2 \times rt \times \int_0^{rmax} \int_0^{2\pi} \frac{G \; ddm}{c \sqrt{c^2 + (h/2)^2}} \frac{(rt - rr \cos(th))}{c} \qquad (2)$$

In digital form this is:

$$v^2 = (pytks)^2 \times rt \times \sum_1^{Nr} 2 \sum_1^{180} \frac{G \; ddm}{\sqrt{c^2 + (h/2)^2}} \times \frac{(rt - rr \cos(th))}{c^2} \qquad (3)$$

where  $c^2 = (rr\ \sin(th))^2 + (rt - rr\ \cos(th))^2$, $pc^2$,  note that c is never zero **

    ddm = mass of the fundamental segment, rho h r dth dr , msuns
    dth  = 1 degree
    G    = gravitational constant, 4.498E-15 $pc^3$ / (msuns / $yr^2$ )
    h    = galaxy equivalent thickness at radius r, pcs
    Nr   = number of rings
  pytks = 9.778E5 (kms/sec)/(pcs/yr)
    r    = radius to centerline of ring, pcs
    rho  = equivalent density, msuns/$pc^3$
    rm   = radius to outer edge of ring, pcs
    rr   = radius to rod used to represent fundamental segment mass, pcs
        = rm -dr / 2 × (rm - 0.575  dr) / (rm - dr / 2)  *
    rt   = radius to test mass, pcs
    th   = (i-1/2) dth, for i = 1 to 180 ,  degs
    v    = orbital speed of test mass at radius rt, kms/sec

* The position of the rod in the fundamental segment was moved slightly from the centroid toward the mid radius of the segment, based on trials to return a given mass distribution from a computed rotation profile.
** Since the rods are at the midangle of each segment, ie at th = 0.5 deg in the first and 179.5 degs in the last segments, c (the distance from test mass to rod) is never zero.

After computing is done in dimensioned form, all output data are made dimensionless by dividing with normalizing parameters:

    accd = acc / acckep , where acckep = GM / $rmax^2$ , the Kepler acceleration at the rim
    md = m(r) / M, where m = mass inside r, and M = total galaxy mass
    hd, rd, rrd, rtd  =  (h, r, rr, rt ) / rmax, where rmax = galaxy rim radius
    rhod = rho / rhoav,  where rhoav = M / (total volume using equivalent segments)
    SMDd = SMD / SMDav,  where SMDav = M / ($\pi$ $rmax^2$ )
    rSMDd = rd × SMDd
    vd, vmd = ( v, vm ) / vkrim , where vkrim = sqr(GM / rmax) , the Kepler speed at the rim

**Reverse Part**
The computing procedure used is:
**1.** input galaxy dimensions, measured speeds at the outer edges of each ring, and arbitrary starting densities for each ring (usually just one density for all rings)
**2.** compute rotation speeds at the outer edges of each ring using (3)
**3.** use speed errors to correct densities of the rings (i = 1 to Nr)
errv = (vm−v) / vmax , f = 0.75 errv , if all errv < 1E−6 then go to 5           (4)
limit abs(f) < 0.5 , rho(j) = (1+f) rho(j−1) for each cycle j
where vm = measured speed and vmax = maximum measured speed
**4.** go to 2 for the next cycle
**5.** make results dimensionless for plots, print or plot results and quit.

Results include total mass, volume, average density, average SMD, Kepler rim speed, maximum computed speed, and the plotted data: md, rhod, rSMDd, vd.

**Examples of graphical results.**

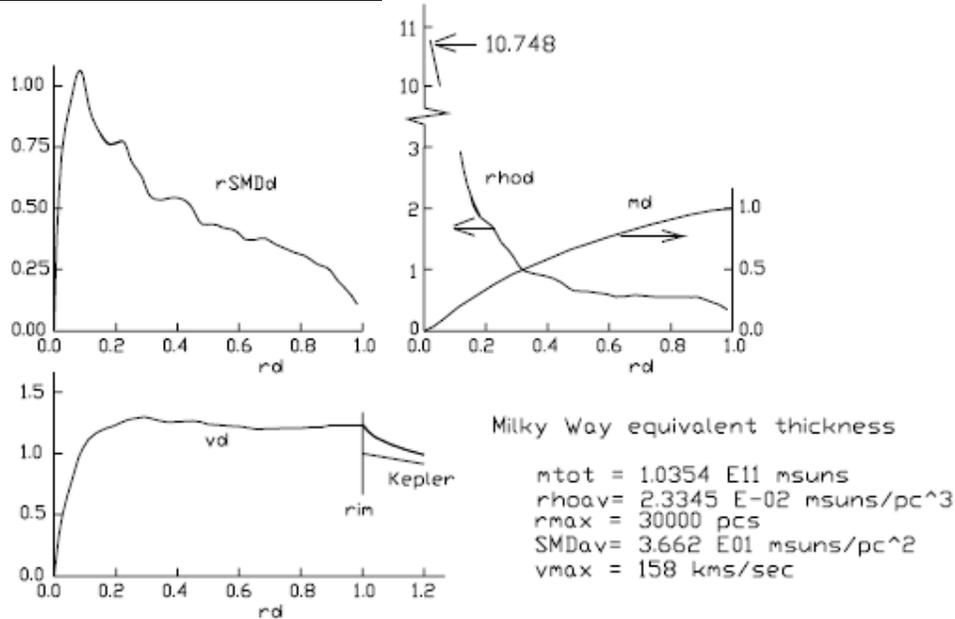

**Figure 3.** Linear scale plots for NGC3198 using rotational speed data from **[3]**.

---

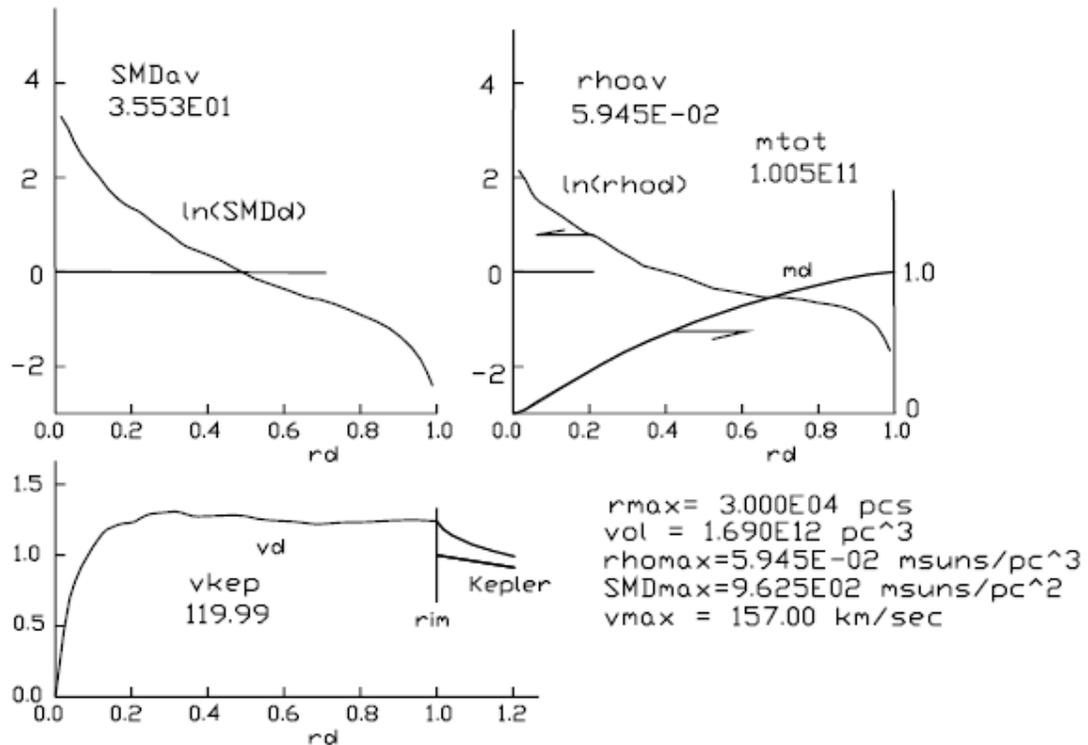

**Figure 4.** Lon (i.e., natural log) scale plots for NGC3198 using rotational speed data from **[3]**. Please notice that, taken at face value, SMD plot here is inconsistent with that in Figure 3.

---

The thickness as a function of radius for NGC3198 is taken to be of the same form as for Milky Way, from a computer-generated side view done by Bab(?)call & Soneira, as compiled by Bok **[4]**. Milky Way rotational speeds, measured by Schmidt & Blitz, are also taken from Bok's compilation **[4]**.

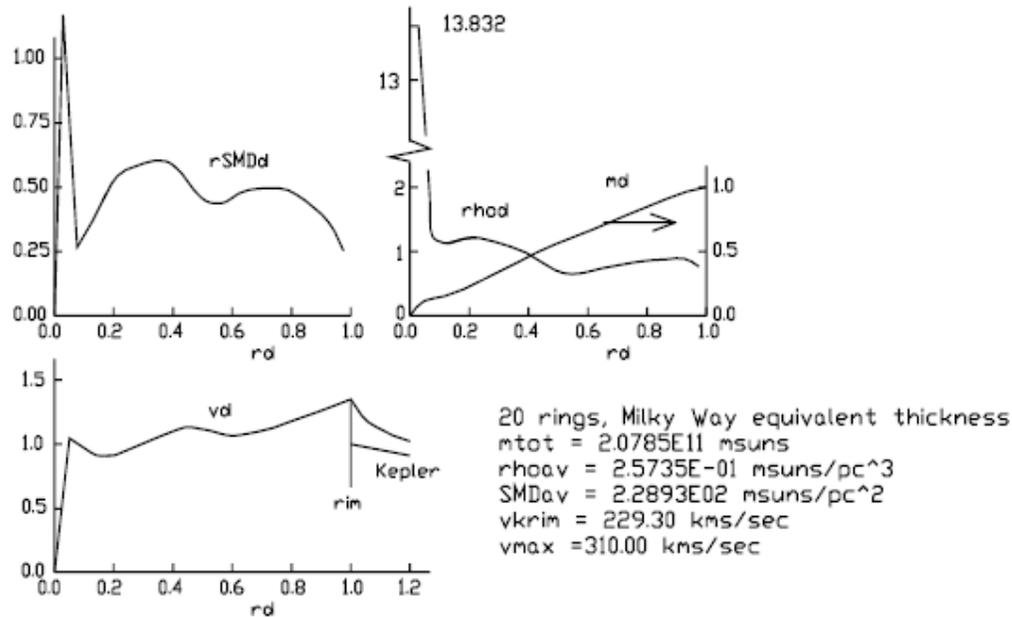

**Figure 5.** Linear scale plots for Milky Way using data from **[4]**.

---------------------------------------------------------------------------

**Conclusions from the two examples.**
Newton's law needs no correction and no dark-matter halos are needed to find galaxy mass distributions from rotation profiles. Given reasonable estimates of galaxy dimensions, including thickness, good values for the mass, SMD, and density distributions are easily found from the rotation profiles. The total masses of galaxies found using dark-matter halos are far too high. Based on reasonable values for dimensions and their rotation profiles, the best values for the total masses of the Milky Way and NGC3198 are 2.079E11 and 1.035E11 msuns respectively.

Nicholson **[2]** presents results for three more examples. A brief summary of some of the parameters for all the five examples is reproduced below.

**Tabular summary from 5 fits**

|          | rmax E-04 | vol E-11 | mtot E-11 | SMDav E-01 | rhoav E02 |
|----------|-----------|----------|-----------|------------|-----------|
| UGC 9133 | 10.250    | 674.000  | 8.274     | 2.261      | 1.228     |
| NGC 3198 | 3.000     | 16.897   | 1.005     | 3.553      | 5.945     |
| Milky Way| 1.700     | 7.556    | 2.053     | 22.669     | 2.717     |
| L M Cloud| 0.900     | 1.121    | 0.089     | 3.476      | 7.890     |
| NGC 6822 | 0.480     | 0.158    | 0.019     | 2.616      | 12.001    |

The units are parsec for the first column, cubic parsec for the second, msuns for the third, msuns/square pc for the fourth & msun/cubic pc for the fifth.

**Outlook.** Researchers interested in rotation curves need to answer certain questions afresh with an open mind. Instead of discounting Kenneth F Nicholson's model just because it is different from the current fashion, it needs to be judged on its own merits. The model calculation needs only modest computing resources. Application to the vast rotation curve and disk thickness data now available should therefore give valuable insights to at least classify galaxies in a sensible scheme. There are two other models available to deal with the data in an axisymmetric approximation **[1, 5]** which also need to be explored without bias from current practice in this area of research. One is a disk galaxy simulation with 50000 particles and the other a matrix inversion calculation with 250000 particles. Comparison of these two and contrast with Nicholson's method promise rich astrophysical dividends.

**Acknowledgments.** I gratefully acknowledge University Grants Commission, New Delhi, India for support. I also thank my family members for comments & help.

[Author's email addresses: dilip.g.banhatti@gmail.com, banhatti@uni-muenster.de]

**Note added in proof:** Kenneth F Nicholson's assumption 4 effectively assumes the same mass-to-light ratio over the full disk, upto its visible extent. The numerical evaluation of the integral transform between the rotation curve and the surface mass distribution SMD(r) thus shows, for the five cases considered, that this works within Newtonian mechanics and gravity. In contrast, R H Sanders (0806.2585 16 June 2008 : From dark matter to MOND) uses a modified Newtonian mechanics towards the same end. Taken at face value, this seems to be a numerical paradox, if the same disk galaxies can be modelled both ways. It will be interesting to examine this contrast further.

-------------------------------------------0x0-------------------------------------------